# Energy landscape picture of supercooled liquids: Application of a generalized random energy model

Masaki Sasai

Graduate School of Human Informatics, Nagoya University, Nagoya 464-8601, Japan

**ABSTRACT**

The thermodynamic and kinetic anomalies of supercooled liquids are analyzed from the perspective of energy landscapes. A mean field model, a generalized random energy model of liquids, is developed which exhibits a dynamical transition of the onset of slow dynamics at $T_0$, alteration of the nature of motion from saddle-to-saddle to minimum-to-minimum motion at $T_c$, and an ideal glass transition at $T_k$. If the energy spectrum of the configurations has a low energy tail, the model also allows a thermodynamic liquid-liquid transition at $T_l$. The liquid-liquid transition of the model is correlated to the kinetic fragile-strong transition accompanied by the anomalous slowing down of motion. Fragility of the system is classified in terms of features of the energy landscape such as ruggedness of the potential energy surface, size of the cooperative motion invoked in transition from one configuration to another, and energy needed to deform the local structure in the cooperative motion. A simple relation is found between diffusion constant, $D$ and the saddle index of the potential energy surface, $f$ as $D \propto f^a$, where $a$ depends on the size of the cooperative motion.

## I. INTRODUCTION

There has been a growing interest in the relation between thermodynamic fluctuations and kinetic slowing down in supercooled liquids[1-6]. The fragility of supercooled liquids has been described with both the thermodynamic and kinetic criteria[1], and calculations of the configurational entropy in computer simulations[3,4] has shown the validity of the Adam-Gibbs relation between entropy and diffusivity[7]. These results suggested intrinsic relationships between thermodynamics and kinetics, which brought our attention to the perspective of the energy landscape of the molecular configurational change[8,9]: The energy landscape should govern both thermodynamic and kinetic behaviors of supercooled liquids and may provide a unified framework of description.

In this paper we develop a simple model of the energy landscape, the random energy model (REM), to see whether it gives a consistent picture of liquids, especially concerning the relation between thermodynamic and kinetic aspects of fragility and the possibility of fragile-strong transitions[10]. In REM, the energy of each configurational state is a random variable representing the rugged energy landscape. Because the statistical features of the energy landscape themselves are the main ingredient (or definition) of a REM, the REM has been used as a most straightforward representation of landscapes in complex systems: REM was first studied by Derrida[11,12] as the simplest model of spin glasses and was applied to problems in protein folding[13-15], random heteropolymers[16,17], and supercooled liquids[5,18-20].

In the simplest REM, however, states are treated as being completely independent of each other and the energy barrier of kinetic movement from state to state should be extremely large, proportional to the system size. This does not explain the quantitative kinetic behaviors of liquids[18]. In order to explain the kinetic properties of liquids, we have to explicitly take into account the correlations among states. In this paper, we utilize the extension made by Derrida[21] and Derrida and Gardner[22], the extension of REM to the generalized random energy model (GREM), to take account of the correlation. It was shown that GREM can fairly well describe the thermodynamic properties of finite dimensional spin glasses[22] and quantitatively describe the kinetics in the lattice model of random heteropolymers[17].

Following the standard arguments on random energy models[11,12], we can see that the GREM of liquids developed in Section II exhibits an ideal thermodynamic glass transition at $T_k$. Thus, the model belongs to the same universality class as models of the random first order transition, where an ideal thermodynamic glass transition underlies the glassy behavior of liquids[23-27]. In Section II the energy spectrum of the correlated energy landscape is analyzed and it will be shown that the model predicts a crossover from saddle-to-saddle motion to minimum-to-minimum motion at $T_c$. Dynamical behavior of the GREM will be analyzed by closely following the method of Wang *et al.*[17], which was originally developed for modeling random heteropolymers, to show the existence of a dynamical transition for the onset of slow dynamics at $T_0$.

For simplicity, distribution of the configurational energy is assumed be Gaussian in Section II. The distribution of energy, however, may be skewed and different from Gaussian in real liquids as will be discussed in Section III. In Section IV, an extension will be made to take account of deviations of the energy distribution from the pure Gaussian by introducing an order parameter which distinguishes the energetically stabilized regions and the entropically favored regions in the liquid. In particular, this extension is necessary to describe fluctuations associated with the hypothesized liquid-liquid transition in water[28] and silica[29,30]. The extended model, the two-component GREM, describes various types of fragility behaviors and allows the liquid-liquid transition at $T_l$ when the energy distribution of configurational states has a low energy tail deviating from the Gaussian. These features will be compared with the molecular dynamics (MD) data of water, silica, and the binary Lenord-Jones (LJ) liquid.

In Section IV and Section V special emphasis will be laid on the anomalous behaviors of liquid water: Both the intense increase in thermodynamic fluctuations and the extreme slowing down in relaxation were observed in the same temperature-pressure region of supercooled liquid water. The thermodynamic anomaly has been hypothetically explained by the liquid-liquid transition[28], while the kinetic anomaly has been considered due to the glass transition[31,32]. The latter interpretation has arisen from the MD observations that the kinetic behaviors of the simulated liquid water are well fitted by the mode coupling theory of the glass transition[32]. We should question whether it is just a coincidence that two different anomalies, thermodynamic and kinetic, take place simultaneously or if there is an intrinsic relationship between the two anomalies. The two-component GREM sheds light on this problem by showing the relation between thermodynamic and kinetic anomalies in the model.

Also important is the simple relation $D \propto f^a$ between the diffusion constant, $D$, and the saddle index of the potential energy surface, $f$, which was found in MD simulations of water[33,34] and silica[35]. This relation directly shows how the kinetics is controlled by the statistical features of the energy landscape. There has been, however, no statistical mechanical explanation of the relation. The two-component GREM gives the condition in which relation $D \propto f^a$ holds and the physical interpretation of the exponent $a$.

## II. GREM OF LIQUIDS

Throughout this paper we consider an ensemble at constant density. As will be explained later in this section, the model gives rise to a distribution in the relaxation time, which is consistent with the picture that the system is decomposed into many mesoscopic regions each of which has different relaxation time[36]. Such heterogeneity in glassy systems has been observed in experiments[37-39] and in MD simulations[40-42]. Though estimation of the size of the mesoscopic region still remains an important problem to be resolved, we focus on one such mesoscopic region and consider the number of molecules in the system, $N$, to be $N \approx 10^2$-$10^5$. We define a 'configurational state' or 'configuration' as a stationary point on the potential energy surface of the $N_d$ dimensional space, where $N_d$ is the total configurational degree of freedom of the system, $N_d = 6N$ in SPC/E model of water[43], for example. The configurational state associated with any phase point is a minimum or a saddle of the potential energy surface, which is obtained by eliminating fast vibrational motions from the snapshot configuration[3,4].

We define $\Omega(E,\rho)dE$ as the number of configurations which have potential energy between $E$ and $E+dE$ at a given density $\rho$. The functional form of $\Omega(E,\rho)$ should depend on the characteristics of each liquid, but in order to explain the physical ingredients of the model, we adopt a simple Gaussian form for $\Omega(E,\rho)$,

$$\Omega(E,\rho)dE = \frac{\Omega_0(\rho)}{\sqrt{2\pi\Delta E(\rho)^2}} \exp\left(-\frac{(E-\overline{E}(\rho))^2}{2\Delta E(\rho)^2}\right) dE, \qquad (2.1)$$

where $\overline{E}$ and $\Delta E^2$ are scaled as $O(N)$, and $dE$ is $O(N^\eta)$ with $\eta < 1$. Deviations of $\Omega(E,\rho)$ from the Gaussian will be taken into account in Section IV.

We assume that kinetics of the system is governed by transition from one configuration to the other. Each configuration corresponds to a saddle or a minimum in the $N_d$ dimensional space, so that the transition should primarily proceed along the direction of small positive curvature or along the direction of negative curvature of the $N_d$ dimensional energy surface. The MD data shows that motions along such directions are cooperative motions consisting of $10^0$-$10^2$ molecules[44,45]. The majority of molecules vibrate along directions of large positive curvature of the energy surface during these motions. Entropy of the system, therefore, is described by a sum of the contribution from saddles or minima, $S_{conf}$, and the contribution from vibrational motions, $S_{vib}$. Such a decomposition of entropy into the configurational part and the vibrational part was discussed in Ref.4. In the present model, the configurational part is described by

$$S_{conf}(E,\rho) = \ln \Omega(E,\rho)dE = S_0(\rho) - \frac{(E-\overline{E}(\rho))^2}{2\Delta E(\rho)^2}, \quad (2.2)$$

where, by retaining the term of order of $N$, $S_0(\rho) \approx \ln\Omega_0(\rho)$. The thermodynamic relation $1/T = dS_{conf}(E,\rho)/dE$ gives an expression of the averaged energy at $T$ as

$$E = \overline{E}(\rho) - \Delta E^2/T. \quad (2.3)$$

Inserting $E$ into $S_{conf}(E,\rho)$, the configuration entropy at temperature $T$ is

$$S_{conf}(T,\rho) = S_0(\rho) - \Delta E^2/2T^2, \quad (2.4)$$

and free energy is $F(T,\rho) = F_{conf}(T,\rho) + F_{vib}$ with

$$F_{conf}(T,\rho) = \overline{E}(\rho) - \Delta E^2/(2T) - TS_0(\rho). \quad (2.5)$$

It should be noted that the larger $\Delta E$ is, the more rapidly entropy decreases, leading to a stronger temperature dependence. In other words, the larger $\Delta E$ is, the more fragile the system is[1].

The model exhibits the glass transition arising from a Kauzmann crisis. At the Kauzmann temperature, $T_k(\rho) = \sqrt{\Delta E^2/(2S_0(\rho))}$, entropy of thermally accessible states becomes less than extensive, $S_{conf}(T,\rho) = 0$ in Eq.(2.4), which should lead to the ideal glass transition. It is expected that before reaching this low temperature limit, the relaxation time should grow large to the macroscopic time scale at temperature $T_g > T_k$. From Eq.(2.4) the amplitude of the specific heat jump at $T_g$ is $\Delta E^2/T_g^2$, which implies the larger amplitude of jump in the more fragile liquid. This feature of REM was compared with experimental data in Ref.5. From Eq.(2.3) we can see that at $T = T_k$ the average energy is $E^- = \overline{E}(\rho) - \sqrt{2S_0(\rho)\Delta E^2}$, which is the representative value of the lowest thermally accessible energy. We use $E^-$ later in this section.

In order to discuss kinetics in the model, not only the energy spectrum but also the relations among configurations have to be explicitly taken into account. The relation between a pair of configurations is described by the similarity parameter, $0 \leq q \leq 1$. When $N_{overlap}$ molecules do not move significantly in the transformation from one configuration to the other, then $q$ is defined by

$$q = N_{overlap}/N. \tag{2.6}$$

In computer simulations $q$ can be measured by detecting saddles from snapshot (instant) configurations of the MD data: Starting from an instant MD configuration, its nearest saddle can be found by minimizing $|\nabla V|^2$, where $V$ is the total potential energy of the instant configuration[46-48]. $N_{overlap}$ and $q$ should be defined by comparing a pair of saddles thus obtained.

Features of the energy landscape are characterized by $S_c(q)$ and $\Delta E_c(q)$. Here, $S_c(q)$ is the entropy of configurations, $S_c(q) = \ln\Omega_c(q)$, where $\Omega_c(q)$ is the number of configurations which have similarity $q$ to a configuration. $\Delta E_c(q)$ is the width of the energy distribution of configurations with similarity $q$ to a configuration. See Fig.1.

Since the probability distribution of $E$ of each configuration, $P(E) = \Omega(E,\rho)/\Omega_0(\rho)$, is Gaussian, probability that a pair of configurations with the mutual similarity $q$ have energy $E_1$ and $E_2$ may also be proportional to a Gaussian,

$$P(E_1, E_2; q) \propto \exp\left(-\frac{(E_1 - E_2)^2}{4\Delta E_c(q)^2}\right). \tag{2.7}$$

$P(E_1, E_2; q)$ has to satisfy the boundary conditions,

$$\lim_{q \to 0} P(E_1, E_2; q) = P(E_1)P(E_2), \tag{2.8}$$

$$\lim_{q \to 1} P(E_1, E_2; q) = P(E_1)\delta(E_1 - E_2). \tag{2.9}$$

The simplest way to satisfy Eq.(2.9) is to put $\Delta E_c(q)^2 = \Delta E^2(1 - q)$. Then, in order to satisfy Eqs.(2.8-9), $P(E_1, E_2; q)$ has the form,

$$P(E_1, E_2; q) = const.\exp\left(-\frac{(E_1 + E_2 - 2\overline{E}(\rho))^2}{4\Delta E(\rho)^2(1+q)}\right)\exp\left(-\frac{(E_1 - E_2)^2}{4\Delta E(\rho)^2(1-q)}\right). \tag{2.10}$$

Eq.(2.10) is the GREM developed by Derrida[21] and by Derrida and Gardner[22]. Though choices other than Eq.(2.10) are possible to satisfy Eqs.(2.8-9), Eq.(2.10) is the simplest and most straightforward expression, so we focus on this model to see the relationships between thermodynamics and kinetics. Models other than Eq.(2.10) will be examined in subsequent publications.

In the following, we suppress the index of $\rho$ from equations when there is no possibility of confusion. $S_c(q)$ should satisfy $S_c(q = 1) = 0$, and $S_c(q = 0) = S_0$. It should be noted that when the functional form of $S_c(q)$ is convex downward or linear, there is a temperature of thermodynamic glass transition at $T_k$, but when $S_c(q)$ is convex upward, the system gradually freezes from a larger to a smaller group of configurations with the occurrence of successive glass transitions as the temperature is lowered[22]. There is no evidence of such successive glass transitions in real liquids. Thus we employ the following piecewise linear functional form with $S' > S_0$ and $S'' < S_0$, which models a downward convex function;

$$S_c(q) = S_0 - q\, S', \quad \text{for} \quad 0 \leq q < q^\ddagger,$$
$$(1-q)\,S'', \quad \text{for} \quad q^\ddagger \leq q \leq 1. \tag{2.11}$$

$S_c(q)$ were estimated to have similar forms in the model of heteropolymers[16]. Here $(1-q)\,S''$ is the entropy increase due to small scale configurational changes such as the microscopic deformation of the molecular cage. $S''$ is smaller than $S_0$ because such configurational change is a correlated motion and the number of reachable configurations with the motion is smaller than the number of uncorrelated randomized configurations. $q^\ddagger$ indicates a crossover from the correlated motion for $q^\ddagger < q$ to the uncorrelated randomized change for $q < q^\ddagger$. We will show later that $q^\ddagger$ measures a characteristic similarity between the starting configuration and the transition state in the cooperative motion from one configuration to another, so that $q^\ddagger$ will play important roles in the model. There are $N_d$ configurations neighboring each configuration, so that the typical similarity measure between neighboring configurations, $q_0$, is given by solving $S_c(q_0) = \ln N_d$ : When $q^\ddagger \leq q_0$, then $q_0 = 1 - \ln N_d / S''$, so that $q_0$ is close to 1 with $1 - q_0 = O((\ln N_d)/N)$.

Notice that in formulating Eq.(2.1), we have defined configurations as stationary points of the potential energy surface including both saddles and minima. Around saddles there are both unstable directions along which the curvature of the potential energy surface is negative and stable directions along which the curvature is positive. We define $f$ to be the average ratio of the number of unstable directions to $N_d$. When one reference configuration has energy $E_1$, the conditional probability that the other neighboring configuration has energy $E_2$ is

$$\xi_{q_0}(E_2; E) = \frac{P(E_1, E_2; q_0)}{P(E_1)}$$

$$= const. \exp\left(-\frac{(E_2 - \overline{E} - q_0(E_1 - \overline{E}))^2}{2(1 - q_0^2)\Delta E^2}\right). \qquad (2.12)$$

Then, the probability $P(f;E)$ that the configuration with energy $E$ has $fN_d$ unstable modes is

$$P(f;E) = {}_{N_d}C_{fN_d} \left(\int_E^\infty dE_2 \xi_{q_0}(E_2;E)\right)^{(1-f)N_d} \left(\int_{-\infty}^E dE_2 \xi_{q_0}(E_2;E)\right)^{fN_d}. \qquad (2.13)$$

The number of configurations which have energy $E$ with the fraction of unstable modes $f$ is $\Omega_f(E) = \Omega(E)P(f;E)$. The average value of $f$ at temperature $T$ is

$$<f> = \frac{\int df \int dE f\, \Omega_f(E) e^{-\beta E}}{\int df \int dE\, \Omega_f(E) e^{-\beta E}}, \qquad (2.14)$$

which is approximated by the saddle-point of the integral, $<f> \approx f^*$, where $f^*$ is obtained by solving $\partial P(f^*;E)/\partial f^* = 0$. From Eqs.(2.12-13), using representative values of $E$ at $T$, $E = \overline{E} - \Delta E^2/T$, we have

$$f^* = \int_0^\infty du \frac{1}{\sqrt{2\pi(1-q_0^2)\Delta E^2}} \exp\left(-\frac{(u + (1-q_0)\beta \Delta E^2)^2}{2(1-q_0^2)\Delta E^2}\right), \qquad (2.15)$$

where $\beta = 1/T$. We focus on the low temperature regime in which $f^*$ is small enough. In this regime,

the integral of Eq.(2.15) can be approximated by an asymptotic expression, leading to

$$f^* \approx \exp\left(-\frac{(1-q_0)\beta^2 \Delta E^2}{2(1+q_0)}\right). \quad (2.16)$$

Eq.(2.16) shows that $f^*$ is a decreasing function of $T$ and at temperature $T = T_c$, the number of unstable modes, $f^* N_d$, becomes smaller than 1,

$$T_c = \sqrt{\frac{(1-q_0)\Delta E^2}{2(1+q_0)\ln N_d}}. \quad (2.17)$$

Notice that when there were no correlation among configurations, then $q_0 = 0$ in the above expressions. In this case of complete randomness in energy, $f^*$ is always very small due to the large value of $\Delta E^2 = O(N)$ and $T_c$ is scaled as $\sqrt{N/\ln N_d}$. In the present model with the correlation among configurations, the small value of $1 - q_0 = O((\ln N_d)/N)$ makes $T_c$ finite, $T_c = O(1)$.

Hereafter, we simply write $f^*$ as $f$. Above $T_c$, $fN_d$ is larger than 1 and the system is usually near saddles and rarely approaches minima. Below $T_c$, on the other hand, $f$ becomes so small that the average number of unstable direction at each configuration is less than 1, which implies that the system predominantly stays around minima instead of saddles. $T_c$ can be regarded as the crossover temperature from the saddle-to-saddle (SS) motion to the minimum-to-minimum (MM) motion. In MD simulations, $f$ can be calculated by the instantaneous normal mode analyses[49-51] which shows that $f$ approaches 0 at the temperature $T_{MCT}$ at which the mode coupling theory predicts divergence of the relaxation time[31,32]. Thus, it should be natural to interpret $T_c$ defined here as the $T_{MCT}$ discussed in Refs.31 and 32. The transformation from the SS type to the MM type motions may imply a change in dynamical features of the liquid from the entropically constrained type to the activation type. Such an interpretation of the SS-MM transformation and its relation to the fragile-strong transformation were discussed by Sasai[52] and Cavagna[53]. In Section IV the physical implication of the SS-MM transformation will be discussed in a more concrete way by using an extended GREM.

Configurations with $f = 0$ are inherent structures. The energy spectrum of the inherent structures is $\Omega_{f=0}(E) = \Omega(E)P(f = 0;E)$. From Eq.(2.13), $P(f = 0;E) = \left(\int_E^\infty dE_2 \xi_{q_0}(E_2;E)\right)^{N_d}$, which can be approximated as $P(f = 0;E) \approx 0$ for $E > E_c$ and $P(f = 0;E) \approx 1$ for $E < E_c$ with $E_c = \overline{E} - \Delta E^2/T_c$. Notice that $\Omega_{f=0}(E)$ is an asymmetric function of $E$ though the low energy side of $\Omega_{f=0}(E)$ is close to a Gaussian form as was observed in the simulation of binary LJ system[3].

From Eq.(2.4) and Eq.(2.16), $S_{conf}(T,\rho)$ and $f$ are related by

$$S_{conf} = \frac{1+q_0}{1-q_0} \ln f + S_0. \qquad (2.18)$$

Here, the coefficient $(1+q_0)/(1-q_0) \approx 2S''/\ln N_d$ is not highly dependent on density when $S''$, which is the rate of the entropy increase with the small scale correlated motion, does not depend on density. The MD simulation of water showed a linear relation between $S_{conf}$ and $\ln f$ and data at different density can be fitted to lines with almost the same gradient[33,34], which suggests that the features of the small scale correlated motion is not highly dependent on density in liquid water[18].

In order to estimate the diffusion constant, we should take into account not only the energy to be surmounted but also the number of paths to reach the next configuration. This estimation can be done by calculating the free energy barrier to be surmounted to reach the next configuration. To derive the expression for the free energy barrier, we closely follow the method of Ref.17 developed for random heteropolymers: We first discuss the hierarchical structure of the energy distribution in the GREM and then use the hierarchical structure to derive the free energy barrier.

First we show that configurations in the GREM are hierarchically organized as shown in Fig.2. Consider two configurations whose mutual similarity is $q$. We write the energies of the two configurations as $E_1 = E^{root} + E_1^>$ and $E_2 = E^{root} + E_2^>$. Then we can see that $E^{root}$, $E_1^>$, and $E_2^>$ are random numbers obeying distribution functions,

$$P^{root}(E^{root}) \sim \exp\left(-\frac{(E^{root} - q\overline{E})^2}{2q\Delta E^2}\right), \tag{2.19}$$

$$P^>(E^>) \sim \exp\left(-\frac{(E^> - (1-q)\overline{E})^2}{2(1-q)\Delta E^2}\right), \tag{2.20}$$

where $E^>$ is either $E_1^>$ or $E_2^>$. Validity of expressions 2.19 and 2.22 can be confirmed by checking whether they satisfy the expected relation $\int dE^{root} P^{root}(E^{root}) P^>(E_1 - E^{root}) = P(E_1)$, and have suitable limits, $P^{root}(E^{root}) \to P(E^{root})$ and $P^>(E^>) \to \delta(E^>)$ when $q \to 1$, $P^{root}(E^{root}) \to \delta(E^{root})$ and $P^>(E^>) \to P(E^>)$ when $q \to 0$. The probability that the two configurations have energy $E_1$ and $E_2$ is

$$P(E_1, E_2; q) = const. \int dE^{root} P^{root}(E^{root}) P^>(E_1 - E^{root}) P^>(E_2 - E^{root}). \tag{2.21}$$

Using Eqs.(2.19-20), we can confirm that Eq.(2.21) is the same as Eq.(2.10). This coincidence implies that GREM is the model in which configurations are organized in a hierarchical way as is shown in Fig.2.

With this hierarchical organization, distribution of $E^{root}$ under the condition that $E_1$ is given is

$$P_q(E^{root} | E_1) = \frac{P^{root}(E^{root}) P_1^>(E_1^>)}{P(E_1)} = const. \exp\left(-\frac{(E^{root} - qE_1)^2}{2q(1-q)\Delta E^2}\right). \tag{2.22}$$

In the limit of $q \to 1$, $P_q(E^{\text{root}}|E_1) \to \delta(E^{\text{root}} - E_1)$, and in the limit of $q \to 0$, $P_q(E^{\text{root}}|E_1) \to \delta(E^{\text{root}})$. The number of configurations which have energy $E$ under the constraint of a given $E^{\text{root}}$ with similarity $q$ to the reference configuration, $\Omega(E:E^{\text{root}},q)$, is $\Omega(E:E^{\text{root}},q) = \Omega_c(q) P^{\text{root}}(E - E^{\text{root}})$. Its logarithm is

$$S(E:E^{\text{root}},q) = \ln \Omega(E:E^{\text{root}},q) = S_c(q) - \frac{(E - E^{\text{root}} - (1-q)\overline{E})^2}{2(1-q)\Delta E^2} . \tag{2.23}$$

From the thermodynamic relation, $1/T = \partial S(E:E^{\text{root}},q)/\partial E$, we have

$$E = E^{\text{root}} + (1-q)\overline{E} - \frac{\Delta E^2}{T}(1-q) . \tag{2.24}$$

Inserting Eq.(2.24) into Eq.(2.23), $S(T:E^{\text{root}},q) = S_c(q) - (1-q)\Delta E^2/2T^2$ Then, the free energy of configurations with energy $E$ and similarity $q$, $F(q) = E - TS(T:E^{\text{root}},q)$, is

$$F(q) = E^{\text{root}} + (1-q)\overline{E} - TS_c(q) - \frac{\Delta E^2}{2T}(1-q) . \tag{2.25}$$

Inserting Eq.(2.25) into Eq.(2.22), the distribution of $F$ is

$$P_q(F) \sim \exp\left(-\frac{1}{2q(1-q)\Delta E^2}(F(q) - (1-q)\overline{E} + TS_c(q) + \frac{\Delta E^2}{2T}(1-q) - qE_1)\right) , \tag{2.26}$$

which has a peak at

$$F(q) = qE_1 + (1-q)\overline{E} - TS_c(q) - \frac{\Delta E^2}{2T}(1-q) \quad . \tag{2.27}$$

Thus, the typical free energy difference between a reference configuration with energy $E_1$ and the ensemble of configurations with similarity $q$ to the reference configuration is $\delta F(E_1,q) = F(q) - E_1$. At sufficiently low temperature, $\delta F(E_1,q)$ has a maximum at $q = q^\ddagger$. The ensemble of configurations with $q^\ddagger$ is, therefore, the transition state of motion from the reference configuration, and $(1- q^\ddagger)N_d$ is the typical size of the cooperative motion arising in the transition from one configuration to the other.

At temperature $T$, the mean escape time from each configuration through one of the transition state configurations is

$$\tau = \tau_0 \int_{E^-}^{E^+} dE_1 P_T(E_1) \exp\left(\delta F(E_1,q^\ddagger)/T\right), \tag{2.28}$$

where $\tau_0$ is the time scale of slow vibrational modes, $P_T(E)$ is the distribution of $E$ at temperature $T$, $P_T(E) \propto \exp(-(E - \overline{E} + \Delta E^2/T)^2 / 2\Delta E^2)$, $E^+$ is defined by $\delta F(E^+, q^\ddagger) = 0$, and $E^- = \overline{E} - \sqrt{2S_0 \Delta E^2}$ is the lowest accessible energy. Eq.(2.28) can be approximated by replacing the integral by the maximum value of the integrant at $E_1^* = \overline{E} - \Delta E^2(2 - q^\ddagger)/T$. Notice that $E_1^*$ is lower than the typical energy at temperature $T$, $\overline{E} - \Delta E^2/T$, showing that the relatively deep minima contribute largely to the determination of the relaxation time constant. The result is summarized as

$$\tau = \tau_0 \exp\left(-S_c(q^\ddagger) + \frac{\Delta E^2}{2T^2}(1-q^\ddagger)(2-q^\ddagger)\right), \quad \text{for } T_k \leq T \leq T_0, \tag{2.29}$$

where $T_0$ is the temperature where the free energy barrier disappears,

$$T_0 = \sqrt{\frac{\Delta E^2 (1-q^\ddagger)(2-q^\ddagger)}{2 S_c(q^\ddagger)}}, \qquad (2.30)$$

For $T > T_0$, $\tau = \tau_0$ and its temperature dependence should obey the Arrhenius formula. By lowering the temperature, the model shows a dynamical transition at $T_0$, which is the onset of the slow kinetics. Below $T_0$, $\tau$ shows a super-Arrhenius type growth following Eq.(2.29). This anomalous growth of $\tau$ is due to the tendency of the system trajectory to survey the lower energy part of the energy landscape as the temperature is lowered, which makes it more difficult to find a small energy barrier to surmount. In the case of $q^\ddagger \approx q_0$, for example, $\tau$ grows so large as $\tau_0 N_d$ at the SS-MM crossover temperature, $T_c$.

Deviation from the Arrhenius behavior can be highlighted by plotting $\ln(\tau/\tau_0)$ vs. $T/T_0$ [54]. From Eqs.(2.29-30),

$$\frac{T}{T_0} \ln \frac{\tau}{\tau_0} = -S_c(q^\ddagger) \left( \frac{T}{T_0} - \frac{T_0}{T} \right), \qquad (2.31)$$

for $T \leq T_0$. Eq.(2.31) is plotted in Fig.3. Though $\Delta E$ in Eq.(2.29) is the apparent density dependent, the normalized representation of Eq.(2.31) fits the data of different density when $S_c(q^\ddagger)$ does not strongly depend on density. Such universality of the curve, $\ln \tau/\tau_0$ vs. $T/T_0$, was found in the MD simulation of binary LJ liquid[55].

Considering $D \approx 1/\tau$, Eq.(2.29) is

$$\ln D/D_0 = S_c - \frac{\Delta E^2}{2T^2}(1-q^{\ddagger})(2-q^{\ddagger}). \tag{2.32}$$

From Eq.(2.16), we have the relation,

$$\ln D/D_0 = S_c(q^{\ddagger}) + \frac{1+q_0}{1-q_0}(1-q^{\ddagger})(2-q^{\ddagger})\ln f, \tag{2.33}$$

for $T \leq T_0$. In this representation, major temperature dependences of $D/D_0$ are renormalized into $f$. When $q^{\ddagger}$ or $q_0$ is not highly dependent on density, a single $D$-$f$ curve of Eq.(2.33) fits the data of different density as was found in the MD simulation of water[33,34]. Notice that the relation holds continuously from $T > T_c$ to $T < T_c$ as was observed in the MD simulation of silica[35]. In Eq.(2.33), the combined effects of a decrease of the number of available configurations and increase of the energy barrier height brings about $D \propto f^a$ with $a \neq 1$ in general. The exponent $a = (1 + q_0)(1 - q^{\ddagger})(2 - q^{\ddagger})/(1 - q_0) \approx 2(1 - q^{\ddagger})/(1 - q_0)$ depends on $q^{\ddagger}$ and $q_0$, that is, the exponent depends on the size of cooperative motions, $(1- q^{\ddagger})N_d$, and the difference between neighboring conformations, $q_0 N$.

Eq.(2.26) shows that the free energy barrier is distributed with the width $\Delta E\sqrt{q^{\ddagger}(1-q^{\ddagger})}$, which implies that the relaxation time constant also distributes with some width. A plausible picture is that liquid is decomposed into many mesoscopic regions, each of which has a different free energy barrier from the other. With this picture, the relaxation of the whole liquid is a superposition of many different relaxation processes, which should result in a non-exponential stretched form of relaxation functions. This mechanism of the stretched relaxation was discussed by Xia and Wolynes by using the random first order transition model[36].

Summarizing this section, the GREM of liquids has salient features which are consistent with experimental observations and MD simulations: The model shows a dynamical transition at $T_0$, at which the temperature dependence of relaxation deviates from the Arrhenius-type behavior and kinetics begins to slow. By lowering temperature, kinetics becomes very slow at $T_c$ and motions are transformed from the saddle-to-saddle type to the minimum-to-minimum type. By further decreasing the temperature, the time constant exceeds the observable scale at $T_g$ showing a jump in specific heat. There underlies the thermodynamic ideal glass transition at $T_k$ below $T_g$. There are simple relations between configuration entropy, $S_{\text{conf}}$, and the saddle index, $f$, and between diffusion constant, $D$, and $f$. The exponent of the latter relation depends on the size of the cooperative motion necessary to surmount the transition state. In this section we used the Gaussian form of the energy spectrum, Eq.(2.1), in order to clarify the physical meaning of the GREM. In the following sections, the model is extended by taking account of deviation from the Gaussian. We will show that fragility can be classified with the extended model of the skewed energy spectrum and discuss the fragile-strong transition and the liquid-liquid transition in the model.

**III TWO-COMPONENT PICTURE OF LIQUID WATER**

There are many experimental and simulation data which suggest that the thermodynamic anomaly of water is due to a 2nd order phase transition between low density amorphous (LDA)-like liquid and high density amorphous (HDA)-like liquid[28,56]. In order to examine this hypothesis in depth, LDA-HDA fluctuations should be analyzed in terms of a suitable order parameter. In the deeply supercooled regime, the local density around each molecule should play the role of the order parameter[57]. In the wider temperature-pressure region, however, fast vibrational motions smear out the difference in local density and another measure is necessary to detect the structural heterogeneity in the system. It was shown that the combined use of the local translational parameter and the local orientational parameter can represent the appropriate structural fluctuation[58]. Another expression of the structural fluctuation is the degree of exclusion of neighboring molecules from the region 3.2Å < $r$ < 3.7Å, where $r$ is the O-O distance from a centered molecule. Shiratani and Sasai[59,60] showed with MD simulations that local structural fluctuations can be quantified by the distribution of neighboring molecules in the region of 3.2Å < $r$ < 3.7Å. A local structure index, $A(\mu, t)$, was defined, which was normalized as $A(\mu, t) \approx 1$ when a central molecule $\mu$ is in the locally structured hydrogen-bond network at a given time instance $t$, and $A(\mu, t) \approx 0$ when $\mu$ is unstructured. Each molecule alternately goes through structured and unstructured periods by distinctive switching transitions and the lifetime of structured periods ranges from several hundred femtoseconds to a few tens of picoseconds at 273K at 1 atm. Structured molecules tend to be located close to each other in space and form clusters[52,60]. The radial distribution function sampled around structured molecules and the one sampled around unstructured molecules mimic the observed radial distribution function of LDA and the one of HDA[61], respectively, so that simulated water is a composite of LDA-like clusters and HDA-like clusters[60].

The growth and collapse processes of these clusters were discussed in Ref.59 and Ref.60. HDA-like molecules are energetically less stabilized than LDA-like molecules[60]. With the instantaneous normal mode analyses, it was shown that the unstable modes are localized on HDA-like molecules, so that the HDA-like molecules are more mobile than the LDA-like molecules[52]. Thus, a 2-component picture which regards liquid water as a composite of two components, LDA-like and HDA-like components, captures many of the important features of structural, energetical, and dynamical heterogeneities in liquid water. The averaged index, $A = <A(\mu, t)>$ which represents the ratio of the number of LDA-like molecules to the total number of molecules should be a convenient order parameter to describe the system.

The heterogeneity of molecules found in the MD data of water may also exist in other liquids. The heterogeneity might play an important role especially in silica in which the liquid-liquid transition is expected[29,30]. When the liquid is decomposed into many heterogeneous parts, then the energy spectrum of configurations should not be a single Gaussian peak but a superposition of many peaks. In the next section, we extend the GREM of the previous section by introducing an order parameter that distinguishes energetically more stabilized structured molecules and less stabilized unstructured molecules.

**IV TWO-COMPONENT GREM**

We represent the concentration of energetically more stabilized molecules as $0 \leq A \leq 1$ and define $\Omega(E,A,\rho)dEdA$ as the number of configurations which have potential energy between $E$ and $E+dE$ with the ratio of stabilized molecules between $A$ and $A+dA$. We assume that $\Omega(E,\rho)$ is a sum over $\Omega(E,A,\rho)dA$,

$$\Omega(E,\rho) = \int dA\, \Omega(E,A,\rho) = \int dA \frac{\Omega_0(A,\rho)}{\sqrt{2\pi \Delta E(\rho)^2}} \exp\left(-\frac{(E-\overline{E}(A,\rho))^2}{2\Delta E(\rho)^2}\right). \quad (4.1)$$

A similar superposition of Gaussians with the order-parameter dependent weight was employed in models of protein folding[13]. Here, by taking account of the mixing entropy, we approximate $S_0(A, \rho) = \ln \Omega_0(A, \rho) = N\alpha(A, \rho) - N(A \ln A + (1-A)\ln(1-A))$. Entropy should be larger in the energetically less stabilized system, so that, $\alpha(A, \rho)$ should be an increasing function of $1-A$. We write $\alpha(A, \rho) = \alpha_0(\rho) + \alpha_1(\rho)(1-A) + \alpha_2(\rho)(1-A)^2$ with $\alpha_0(\rho), \alpha_1(\rho)$, and $\alpha_2(\rho) > 0$. $\overline{E}(A,\rho)$ should be a decreasing function of $A$, such that $\overline{E}(A,\rho) = N(e_0(\rho) + e_1(\rho)A + e_2(\rho)A^2)$ with $e_1(\rho)$ and $e_2(\rho) < 0$. Dependence of $\Delta E(\rho)$ on $A$ is neglected for simplicity.

When $A$ is constrained to be a constant, the same expression as Eq.(2.5) is obtained for $F_{\text{conf}}$,

$$F_{\text{conf}}(T, A, \rho) = \overline{E}(A,\rho) - \Delta E^2/(2T) - TS_0(A,\rho). \quad (4.2)$$

The most probable value of $A = A^*(T, \rho)$ is evaluated from the variation, $\partial F(T, A^*, \rho)/\partial A^* = 0$. We assume that $F_{\text{vib}}$ does not strongly depend on $A$, so that $A^*$ is approximately determined by $\partial F_{\text{conf}}(T, A^*, \rho)/\partial A^* = 0$. The Kauzmann temperature is written as

$$T_k(A^*, \rho) = \sqrt{\Delta E^2 /(2S_0(A^*,\rho))}, \qquad (4.3)$$

where entropy $S_{\text{conf}}(T, A^*, \rho)$ diminishes to be 0, leading to the ideal glass transition. Since $S_0(A^*, \rho)$ is smaller in the large $A^*$ state, the model predicts that $T_k$ is higher when $A^*$ is larger.

$F_{\text{conf}}(T, A, \rho)$ is shown in Fig.4 as a function of $A$. When dependence of energy on $A$ is small with small values of $|e_1|$ and $|e_2|$, then the free energy minimum is always at small $A^*$ as shown in Fig.4a. When $|e_1|$ and $|e_2|$ are large, on the other hand, the minimum is at small $A^*$ at high temperature, but shifts to the larger $A^*$ as $T$ is lowered (Fig.4b-d). The model permits both a discontinuous transition (Fig4.b-c) and a continuous transformation (Fig.4d). When $\alpha_2$ and $|e_2|$ are large, the system exhibits a discontinuous transition as shown in Fig.4c at temperature $T_l \approx (\overline{E}(A^*) - N(e_0 + e_1 + e_2)/(S_0(A^*, \rho) - N\alpha_0)$. When $\alpha_2$ or $|e_2|$ is small, on the other hand, $F_{\text{conf}}(T, A, \rho)$ has a single minimum at $A = A^*(T, \rho)$ and $A^*(T, \rho)$ continuously increases from the small $A$ state with $A^*(T, \rho) < 1$ to the large $A$ state as $T$ decreases. Thus, a discontinuous liquid-liquid transition is possible when the energy spectrum has large dependences on $A$. Parameters in the energy spectrum such as $\alpha_2$ and $e_2$ should depend on density, so that it would be natural to assume that the discontinuous liquid-liquid transition is possible only within a certain range of density ( or pressure ). In this case, the transition line may be terminated at a critical point. Around this critical point, the model should exhibit an anomalous increase in thermodynamic response functions as was observed in supercooled liquid water[56].

Correlations between configurations are taken into account by extending Eqs.(2.19-20) in the following way: Consider two configurations denoted by $(E_1, A_1)$ and $(E_2, A_2)$ whose mutual similarity is $q$. $|A_1 - A_2|$ should be smaller than $1 - q$ by definition. $E_1$ and $E_2$ are expressed as $E_1 = E^{\text{root}} + E_1^>$ and $E_2 = E^{\text{root}} + E_2^>$, where $E^{\text{root}}$ and $E_1^>$ are random numbers obeying distribution functions,

$$P^{\text{root}}(E^{\text{root}}) \sim \exp\left(-\frac{(E^{\text{root}} - q\overline{E}(A_1))^2}{2q\Delta E^2}\right), \tag{4.4}$$

$$P_1^>(E_1^>) \sim \exp\left(-\frac{(E_1^> - (1-q)\overline{E}(A_1))^2}{2(1-q)\Delta E^2}\right), \tag{4.5}$$

and the distribution of $E_2^>$ is expressed by comparing it to $\overline{E}(A_1)$ as

$$P_2^>(E_2^>) \sim \exp\left(-\frac{(E_2^> - (1-q)\overline{E}(A_1) - \delta\overline{E})^2}{2(1-q)\Delta E^2}\right), \tag{4.6}$$

where $\delta\overline{E} = \overline{E}(A_2) - \overline{E}(A_1)$. We can confirm that Eqs.(4.4-6) satisfy desirable limits: In the limit of $q \to 1$, $P_1^>(E_1^>) \to \delta(E_1^>)$ and $P_2^>(E_2^>) \to \delta(E_2^>)$ and in the limit of $q \to 0$, $P_1^>(E_1^>) \to P(E_1^>, A_1)$ and $P_2^>(E_2^>) \to P(E_2^>, A_2)$. When $\delta\overline{E} = 0$, Eqs.(4.4-6) are reduced to GREM relations of Eqs.(2.19-20). From Eqs.(4.4-6), $P(E_1, E_2; q) = const.\int dE^{\text{root}} P^{\text{root}}(E^{\text{root}})P_1^>(E_1 - E^{\text{root}})P_2^>(E_2 - E^{\text{root}})$. Then, following the derivation from Eq.(2.12) to Eq.(2.16) and by writing $f^*$ as $f$ for simplicity, the saddle index at temperature $T = 1/\beta$ is

$$f \approx \exp\left(-\frac{((1-q_0)\beta\Delta E^2 + \delta\overline{E}_0)^2}{2(1-q_0^2)\Delta E^2}\right), \tag{4.7}$$

where $\delta\overline{E}_0 = \overline{E}(A^* - (1-q_0)) - \overline{E}(A^*)$. The SS-MM crossover temperature, $T_c$, is obtained by

solving $fN_d = 1$ as

$$T_c = \left( \sqrt{\frac{2(1+q_0)\ln N_d}{(1-q_0)\Delta E^2}} - \frac{\delta \overline{E}_0}{(1-q_0)\Delta E^2} \right)^{-1}. \quad (4.8)$$

A measure of fragility of the liquid should be given by $0 < T_k/T_c < 1$,

$$\frac{T_k}{T_c} = \sqrt{\frac{(1+q_0)\ln N_d}{(1-q_0)S_0}} \left( 1 - \frac{\delta \overline{E}_0}{\sqrt{4(1-q_0)\Delta E^2 \ln N_d}} \right), \quad (4.9)$$

which shows that the liquid is more fragile when $\delta \overline{E}_0$ is smaller, *i.e.* the energy dependence on $A$ is smaller or when $\Delta E$ is larger, *i.e.* the energy landscape is more rugged.

The energy spectrum of the inherent structures is $\Omega_{f=0}(E) = \int dA\, \Omega(E, A)P(0: E, A)$. In Fig.5, the entropy of the inherent structures, $S_{inh}(E) = \ln \Omega_{f=0}(E)$, is drawn together with the entropy of configurations, $S_{conf}(E) = \ln \int dA\, \Omega(E, A)$, and entropy of the large $A$ state, $S_{largeA}(E) = \ln \Omega(E, A>0.85)$. There is no liquid-liquid transition when the form of the energy spectrum is close to Gaussian (Fig.5a). When the spectrum has a low energy tail as shown in Fig.5c, there is a liquid-liquid transition at temperature $T_l$.

Since $S_{largeA}(E)$ and $S_{inh}(E)$ have a large overlap, it is expected that the liquid-liquid transition and the SS-MM transition should take place simultaneously: $T_l \approx T_c$. This point is more clearly displayed in the plot of Fig.6. $T_c$ in Eq.(4.8) depends on $A^*$ through $\delta \overline{E}_0$, and $A^*$ is a function of $T$. Eq.(4.8), therefore, represents a self-consistent relation for the temperature: The SS-MM crossover temperature is obtained by solving $T_c(T) = T$. In Fig.6, $T^{-1} - T_c(T)^{-1}$ is plotted together with $A^*$ in the case where a liquid-liquid transition exists at the low temperature regime. Because $T_c(T)$ shows a jump at $T_l$ where $A^*$ shows a jump, there is a good chance that $T^{-1} - T_c(T)^{-1}$ crosses 0 at $T_l$. In Fig.6a, $T^{-1} - T_c(T)^{-1}$ crosses 0 at $T_l$, which implies $T_l = T_c$. In Fig6b with a smaller $N_d$ than in Fig.6a, $T_c$ is slightly above $T_l$ but in general, $T_l$ and $T_c$ are correlated to each other. We can say, therefore, that the liquid-liquid transition marked by a change of $A^*$ is associated with the SS-MM transition of diminishing $f$.

At $T < T_0$, the rate limiting step for motion from a given configuration with energy $E_1$ and the order parameter $A^*$ is to surmount the free energy barrier at the transition state configurations, configurations with similarity $q^{\ddagger}$ and the order parameter $A^{\ddagger}$. The representative value of free energy at the transition state is

$$F(q^{\ddagger}, A^{\ddagger}) = q^{\ddagger} E_1 + (1 - q^{\ddagger}) \overline{E}(A^*) + \delta \overline{E}^{\ddagger} - T S_c(q^{\ddagger}, A^{\ddagger}) - \frac{\Delta E^2}{2T}(1 - q^{\ddagger}), \qquad (4.10)$$

where $\delta \overline{E}^{\ddagger} = \overline{E}(A^{\ddagger}) - \overline{E}(A^*)$ and $A^{\ddagger}$ should be chosen to minimize free energy under the restriction, $|A^{\ddagger} - A^*| < 1 - q^{\ddagger}$. The free energy barrier to be surmounted is $\delta F = F(q^{\ddagger}, A^{\ddagger}) - E_1$, which determines the relaxation time. Averaging over $E_1$, the representative value for the relaxation time is

$$\tau = \tau_0 \exp\left(-S_c(q^{\ddagger}, A^{\ddagger}) + \frac{\Delta E^2}{2T^2}(1-q^{\ddagger})(2-q^{\ddagger}) + \frac{\delta \overline{E}^{\ddagger}}{T}\right), \qquad \text{for } T_k \leq T \leq T_0. \qquad (4.11)$$

In Eqs.(4.10) and (4.11), $S_c(q, A)$ is given by the extension of Eq.(2.11),

$$S_c(q, A) = S_0(A) - q S'(A), \quad \text{for} \quad 0 \leq q < q^{\ddagger},$$
$$(1-q) S''(A), \quad \text{for} \quad q^{\ddagger} \leq q < 1, \qquad (4.12)$$

with $S''(A) < S_0(A)$. The onset temperature of slow dynamics, $T_0$, is obtained by solving $\tau = \tau_0$ in Eq.(4.11) as

$$T_0 = \frac{\delta \overline{E}^{\ddagger}}{2S_c(q^{\ddagger}, A^{\ddagger})} + \sqrt{\left(\frac{\delta \overline{E}^{\ddagger}}{2S_c(q^{\ddagger}, A^{\ddagger})}\right)^2 + \frac{\Delta E^2 (1-q^{\ddagger})(2-q^{\ddagger})}{2S_c(q^{\ddagger}, A^{\ddagger})}} \ . \qquad (4.13)$$

From Eq.(4.11), we can see that $\tau$ shows the super-Arrhenius behavior for $T < T_0$ when $\Delta E^2(1 - q^{\ddagger})(2 - q^{\ddagger})$ is more significant than the term of $\delta \overline{E}^{\ddagger}$, and $\tau$ shows the Arrhenius-type temperature dependence when $\delta \overline{E}^{\ddagger}$ is important. The system is "fragile" in the former case and "strong" in the latter case. Eq.(4.11), therefore, represents both fragile and strong behaviors depending on parameters to express features of the energy landscape, $\Delta E$, $q^{\ddagger}$, and $\delta \overline{E}^{\ddagger}$. In Fig.7 various fragility behaviors are exemplified by the "Angell plot", the plot of $\ln \tau/\tau_0$ vs. $T_g/T$. When $|e_1|$ and $|e_2|$ are small, $\delta \overline{E}^{\ddagger}$ is less important, so that the system is fragile as in Fig.7a but when $|e_1|$ and $|e_2|$ are significant, the system is stronger as in Fig.7b.

In Fig.7c, we can find that the system alters its fragility when the liquid-liquid transition intervenes. $A^*$ is small for $T > T_l$ and large for $T < T_l$, so that $\delta \overline{E}^{\ddagger}$ is less important for $T > T_l$ than for $T < T_l$. In this way, the system shows the transition at $T = T_l$, from the fragile liquid for $T > T_l$ to the strong liquid for $T < T_l$. As was explained in the discussion for Fig.6, $T_c = T_l$. Therefore, three transitions, the thermodynamic liquid-liquid transition shown in Fig.4c, the SS-MM transition explained in Fig.6a, and the fragile-strong transition shown in Fig.7c coincide at the same temperature. By varying parameters $e_2$ or $\alpha_2$, the line of the liquid-liquid transition is terminated at a critical point. Around the critical point, $A^*$ shows a continuous but sudden change. In this super-critical case, the system continuously transforms from fragile to strong by lowering the temperature as shown in Fig.7d. Here, three transformations, the liquid-liquid transformation, the SS-MM transformation, and the fragile-strong transformation take place at near, but not necessarily coinciding, temperatures. As the critical point is approached, however, the temperatures of these three transformations come closer. This convergence of temperatures might provide a clue in the search for the location of the critical point in the phase diagram. Existence of the fragile-strong transformation in water was suggested from its observed[10,62.63] and simulated[63] data and also in silica from the simulations[29,30]. There remains, however, a debate on the existence of the fragile-strong transformation in water[64]. The present model showed that as the thermodynamic critical point is approached, the sharper the fragile-strong transition becomes. This correlation should provide an important clue to resolve the debate when the experimental data becomes available for the wider temperature-pressure regions.

From Eqs.(4.7) and (4.11) and $D \approx 1/\tau$, we have the relation for $T \leq T_0$,

$$\ln D/D_0 = S_c + \frac{(1-q^{\ddagger})(2-q^{\ddagger})}{1-q_0}\left((1+q_0)\ln f + \frac{1}{T}\left(\delta\overline{E}_0 - \frac{1-q_0}{(1-q^{\ddagger})(2-q^{\ddagger})}\delta\overline{E}^{\ddagger}\right) + \frac{\delta\overline{E}_0^{\,2}}{2(1-q_0)\Delta E^2}\right).$$

(4.14)

When $1 - q^{\ddagger} = 1 - q_0 = O(\ln N_d/N)$, that is, when the size of the cooperative motion is as small as the rearrangement of the microscopic molecular cage, then, Eq.(4.14) is reduced to a simpler form of $D \propto f^2$, which is the relation found in MD simulation of water[33,34]. When $f$ is so small, $\ln f$ becomes more significant in Eq.(4.14) than $\delta\overline{E}_0$ or $\delta\overline{E}^{\ddagger}$. Then, Eq.(4.14) results in a simple form again,

$$\ln D/D_0 = S_c + \frac{1+q_0}{1-q_0}(1-q^{\ddagger})(2-q^{\ddagger})\ln f \,. \qquad (4.15)$$

This relation was found in the MD simulation of silica[35], where the exponent was 1.3. The relation $(1 + q_0)(1 - q^{\ddagger})(2 - q^{\ddagger})/(1 - q_0) = 1.3$ should be checked with MD simulation by calculating $q^{\ddagger}$ and $q_0$ in silica.

In similar ways the relation between $\ln \tau/\tau_0$ and $T/T_0$ and the relation between $S_{\text{conf}}$ and $f$ are reduced to simple forms in the large $\tau$ limit or in the small $f$ limit,

$$\frac{T}{T_0} \ln \frac{\tau}{\tau_0} = -S_c(q^{\ddagger}, A^{\ddagger})\left(\frac{T}{T_0} - \frac{T_0}{T}\right), \tag{4.16}$$

and

$$S_{\text{conf}}(A^*) = \frac{1+q_0}{1-q_0} \ln f + S_0(A^*). \tag{4.17}$$

By using a REM without correlation, relations $D \propto f$ and $\exp(S - S_0) \propto f$ were derived[18]. In the present work, it was shown that correlations among configurations modify the exponents as $D \propto f^a$ and $\exp(S - S_0) \propto f^{b/(1-q_0)}$ with $b = (1 + q_0) \approx 2$ and $a = b(1 - q^{\ddagger})(2 - q^{\ddagger})/(1-q_0) \approx b(1 - q^{\ddagger})/(1 - q_0)$. Further examination of $q^{\ddagger}$ and $q_0$ in MD simulations will clarify whether the energy landscape picture proposed here is consistent.

## V SUMMARY AND DISCUSSION

In spite of its simplicity, the 2-component GREM developed in this paper captures many features of observed and simulated data of supercooled liquids. When the energy spectrum has a low energy tail deviating from the Gaussian, a discontinuous liquid-liquid transition is possible at temperature, $T_l$. When this transition line is terminated at a critical point, thermodynamic response functions should show a strong increase around the critical point. The present model showed that the liquid-liquid transformation around the critical point should be correlated to the SS-MM transformation, which is associated with the large increase of the relaxation time constants. In supercooled liquid water, simultaneous increase of both specific heat and viscosity has been observed[28,56]. There has been, however, no clear understanding on whether the simultaneous increase of thermodynamic fluctuations and relaxation constants is simply a coincidence or if there is a common mechanism underlying both anomalies. The present model proposed the scenario that the energy landscape traversed by dynamic trajectories changes at the liquid-liquid transition point, which necessarily gives rise to the SS-MM transition. This scenario should be further examined by both MD simulations and experiments, which should then give a clue to construct the phase diagram of supercooled liquid water.

Types of various fragility are classified in terms of ruggedness of the energy surface, $\Delta E$, size of the cooperative motion, $(1-q^{\ddagger})N_d$, and energy needed to deform the local structure upon cooperative motion, $\delta \overline{E}^{\ddagger}$. When the liquid-liquid transition intervenes, kinetic features are altered from fragile to strong ones at the liquid-liquid transition point. The correlation between the liquid-liquid transition and the fragile-strong transition found in the present model should help to resolve the debate on whether the fragile-strong transition really exists in liquids.

It has been shown by MD simulations of water[33,34] and silica[35] that there are simple relations between $S_{\text{conf}}$ and $f$, and between $D$ and $f$. These relations strongly support the idea that the energy landscape governs both the thermodynamics and kinetic behaviors of these liquids. The relation $S_{\text{conf}} - S_0 \propto \ln f$ was derived by using REM by Keyes[18], but no statistical mechanical explanation has been given to the relation between $D$ and $f$ because of the lack of a suitable model of the correlated energy landscape. The present model showed that $D \propto f^a$ and that the exponent $a$ depends on the size of the cooperative motion from one configuration to another. The mechanism to determine $a$ proposed in the present model should be further checked by MD simulations.

The model was defined by functions which characterize the energy landscape such as $S_c(q, A)$, $\Delta E_c(q)$, and $\overline{E}(A, \rho)$. Notice that the model does not start from a microscopic Hamiltonian, so that it does not directly give results displaying the effects of a specific interaction or a specific geometrical constraint. Rather, it gives results once the functions which characterize the energy landscape are given. The model, therefore, connects knowledge on the energy landscape and thermodynamic and kinetic features in a systematic way, which will aid in the analysis of observed and simulated data to find the fundamental mechanism of anomalies of supercooled liquids.

In the present treatment, there are many points left to be improved. For example, we used simple forms of functions for $S_c(q, A)$, $\Delta E_c(q)$, and $\overline{E}(A, \rho)$ in order to simplify the argument. These functions should be replaced by more realistic ones in order to compare the model with experiments in a more quantitative way. A most drastic approximation used here was to make $q^{\ddagger}$ independent of $A$. When dependence of $q^{\ddagger}$ on $A$ is taken into account, temperature dependence on the size of the cooperative motion should be more reasonably treated. Further quantitative comparison to observed and simulated data will be possible through these improvements.

**ACKNOWLEDGMENTS**

The author thanks Prof. T. Keyes for valuable comments. This work was supported by Japan Science and Technology Corporation (ACT-JST), Japan-US cooperative research program of Japan Society of Promotion of Science, and Ministry of Education, Science, Sports, and Culture, Japan.

# FIGURE CAPTIONS

Figure 1

The energy landscape is characterized by entropy, $S_c(q)$, and the width of energy distribution, $\Delta E_c(q)$, of configurations which have similarity $q$ (dissimilarity $1 - q$) to an arbitrarily chosen configuration.

Figure 2

In GREM, configurations are hierarchically organized with an ultrametric relation: Energy $E_1$ and energy $E_2$ have a common root of energy $E^{\text{root}}$.

Figure 3

Dynamical transition of the onset of slow dynamics at $T_0$. Below $T_0$, $\tau$ increases in a super-Arrhenius way. The model predicts that the data of different density should fall on a single line when the transition state is not highly dependent on density. Here, $q^{\ddagger}$ was set to be $q^{\ddagger} = q_0$, so that $S_c(q^{\ddagger}) = S_c(q_0) = \ln N_d$ with $N_d = 1000$.

Figure 4

Dependence of the constrained free energy, $F_{\text{conf}}(T, A)/N$, on $A$ at various $T$. (a) $F_{\text{conf}}(T, A)$ has a single minimum at small $A$. $e_1 = -0.1$, $e_2 = -0.2$, $\alpha_1 = 1.0$, and $\alpha_2 = 1.0$. (b) $F_{\text{conf}}(T, A)$ has double minima at small $A$ and large $A$ and discontinuous transition takes place between them as $T$ is altered. Free energy barrier of the transition, however, is very small. $e_1 = -1.0$, $e_2 = -1.0$, $\alpha_1 = 1.0$, and $\alpha_2 = 1.0$. (c) A clear discontinuous transition takes place between the small $A$ state and the large $A$ state. $e_1 = -1.0$, $e_2 = -1.5$, $\alpha_1 = 0.5$, and $\alpha_2 = 1.2$. (d) $F(T, A)$ has a single minimum and continuous transformation is brought about by changing $T$. $e_1 = -1.0$, $e_2 = -1.5$, $\alpha_1 = 0.5$, and $\alpha_2 = 0.85$. $\Delta E = 1.0$ and $\alpha_0 = 1.0$ for all cases of (a)-(d).

Figure 5

Energy dependence of entropy of total configurations, $S_{conf}(E)/N$, is plotted with a thick line, entropy of inherent structures, $S_{inh}(E)/N$, with a thin line, and entropy of configurations with $A > 0.85$, $S_{large\,A}(E)/N$, with a dashed line. The same parameter sets as in Fig.4a, b, and c are used for (a), (b), and (c), respectively. $N=200$ and $N_d = 1000$ for (a) through (c). $(1-q_0)N=30$ in (a) and (b), and $(1-q_0)N=6$ in (c).

Figure 6

The liquid-liquid transition takes place at $T_l^{-1} = 0.68$ with a discrete jump of $A$ (real line). The self consistent relation, $T = T_c$, holds at temperature where the curve of $T^{-1} - T_c^{-1}(T)$ (dashed line) crosses 0. (a) $T_c^{-1} = T_l^{-1}$ or (b) $T_c^{-1} = 0.61 < T_l^{-1} = 0.68$. The same parameter set as in Fig.4c is used with $(1-q_0)N = 6$ for both (a) and (b). $N_d = 50000$ in (a) and $N_d = 1000$ in (b).

Figure 7

Various types of fragility behaviors are shown by plotting $\ln \tau/\tau_0$ as a function of $T_g/T$. $T_g$ is defined by the temperature at which $\ln \tau/\tau_0 = 25$. $\ln \tau/\tau_0$ is more separated from the dashed line of the Arrhenius relation in (a) than in (b). (c) A discontinuous liquid-liquid transition is associated with the fragile-strong transition. (d) A continuous liquid-liquid transformation is associated with the fragile-strong transformation. Same parameters as in Fig.4a-d are used for (a)-(d), respectively. $(1-q_0)N = 30$ in (a) and (b), $(1-q_0)N = 6$ in (c), and $(1-q_0)N = 8$ in (d). $q^{\ddagger} = q_0$, $N_d = 50000$, and $S''(A^{\ddagger}) = 10^{-2} S_0(A^{\ddagger})$ for all cases. Characteristic temperatures are obtained by solving Eqs.(4.3), (4.8), and (4.13) self-consistently as (a) $T_g/T_0 = 0.13$, $T_g/T_c = 0.90$, and $T_g/T_k = 2.05$ with $T_g = 0.830$. (b) $T_g/T_0 = 0.06$, $T_g/T_c = 0.17$, and $T_g/T_k = 4.28$ with $T_g = 1.739$ and $T_g/T_l = 1.74$. (c) $T_g/T_0 = 0.03$, $T_g/T_c = 0.71$, and $T_g/T_k = 1.60$ with $T_g = 1.040$ and $T_g/T_l = 0.71$. (d) $T_g/T_0 = 0.03$, $T_g/T_c = 0.57$, and $T_g/T_k = 2.16$ with $T_g = 1.300$.

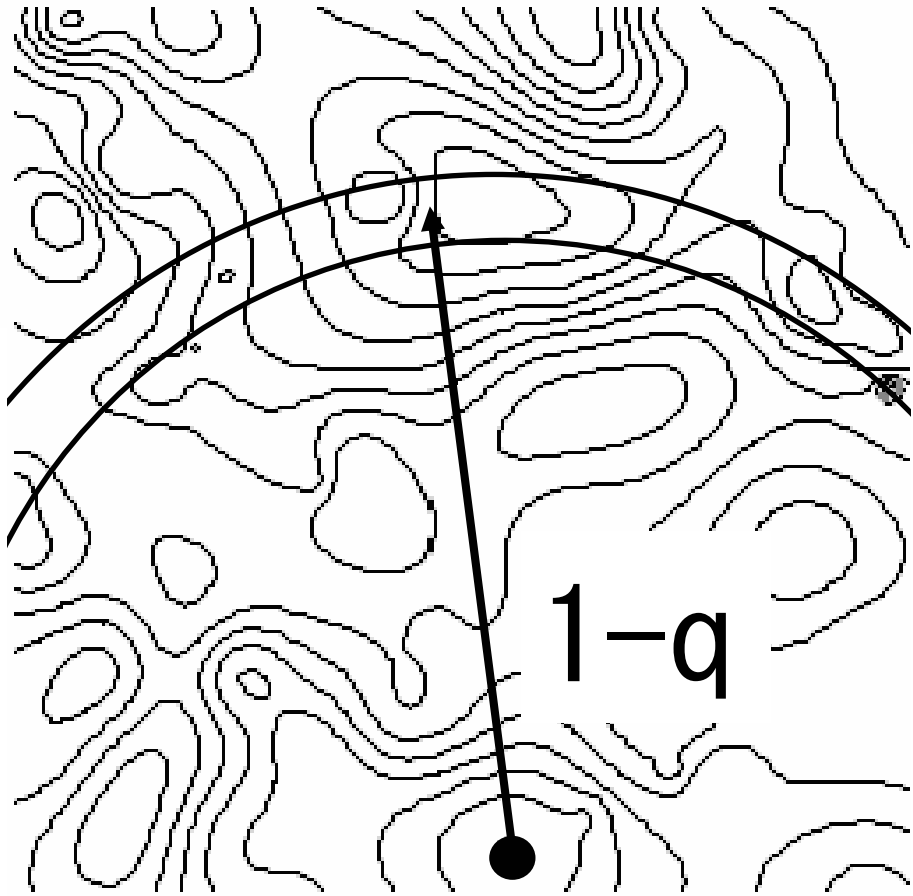

Fig.1 Sasai

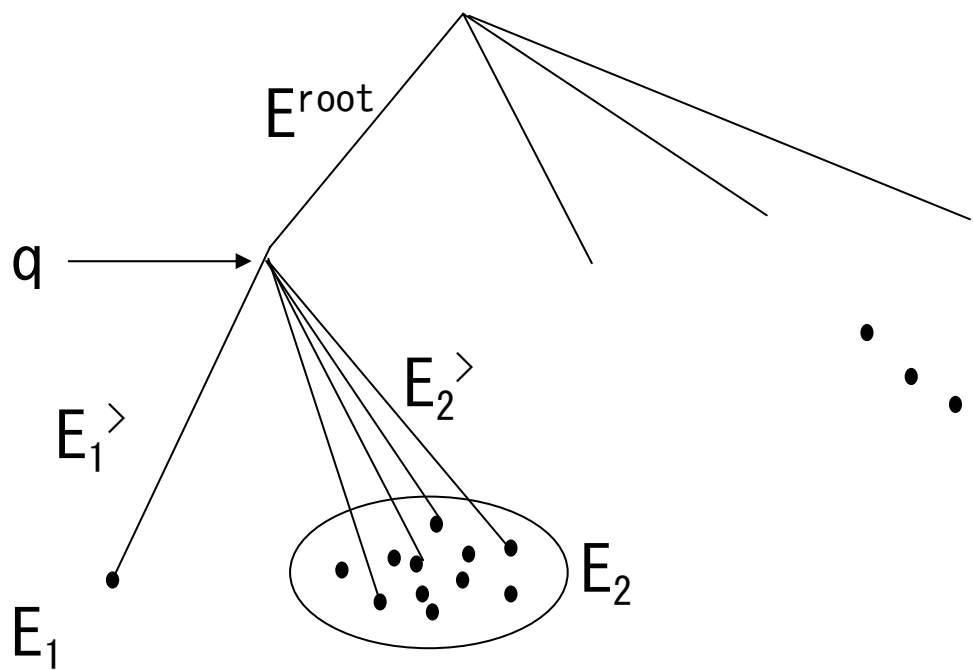

Fig.2 Sasai

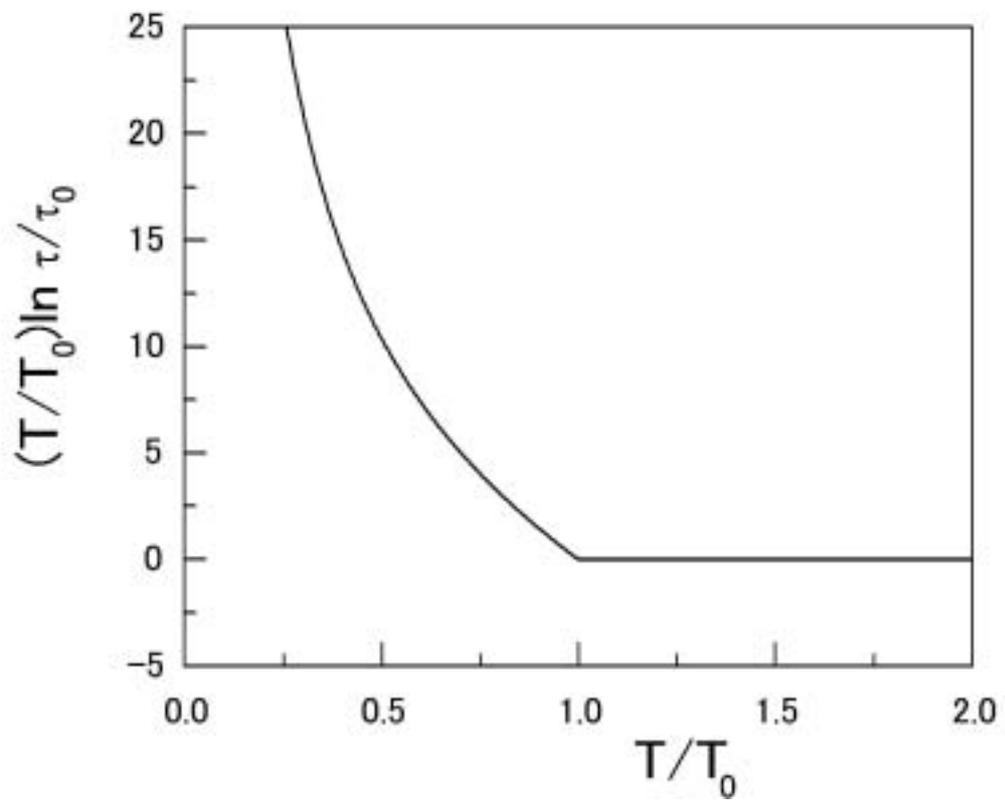

Fig.3 Sasai

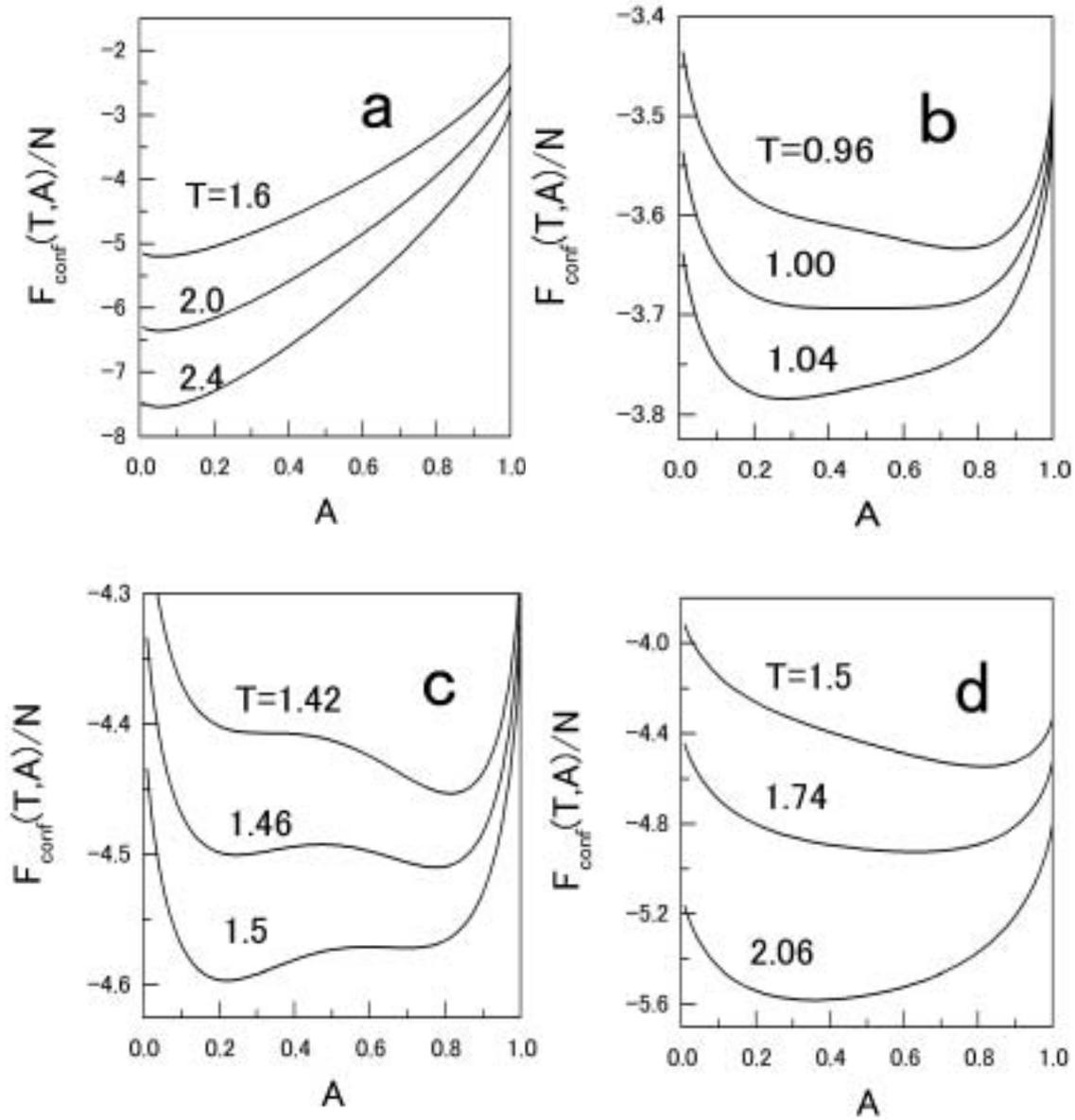

Fig.4 Sasai

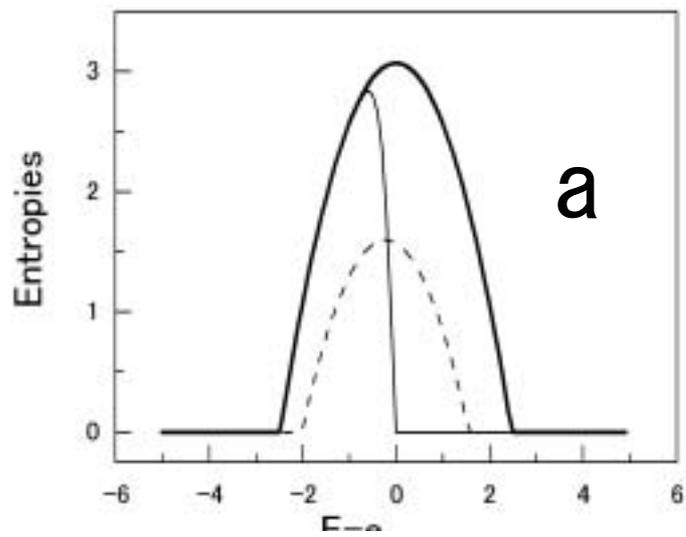

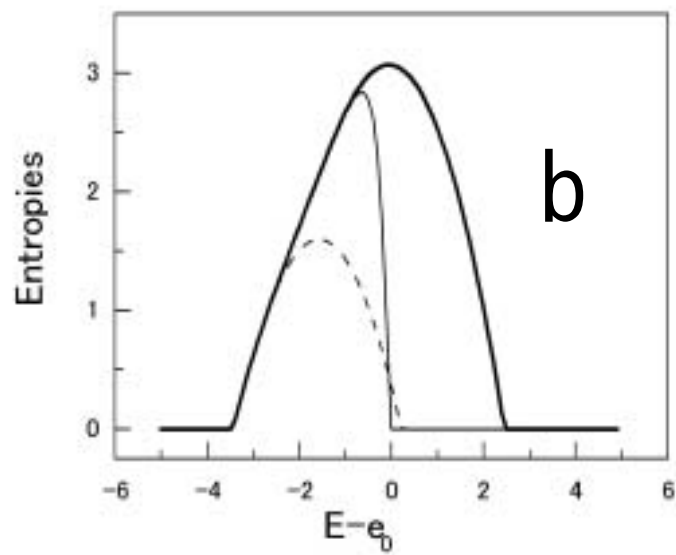

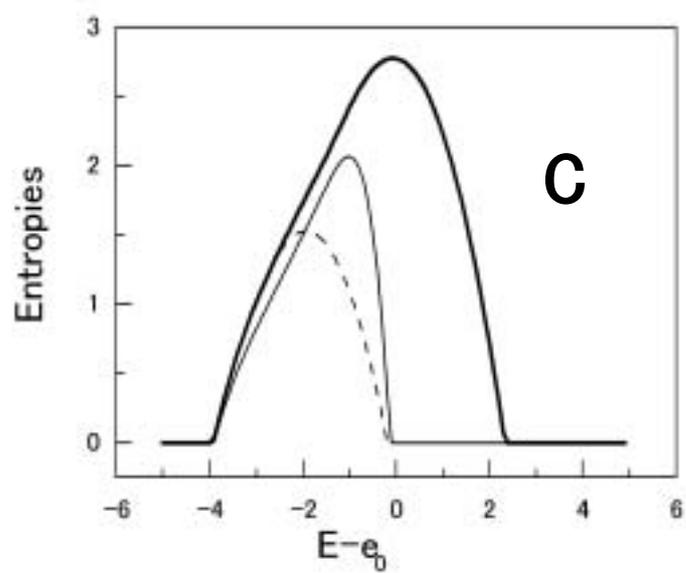

Fig.5 Sasai

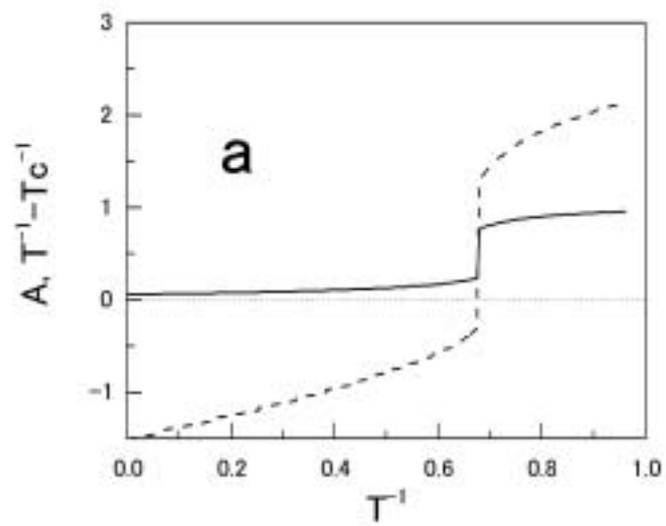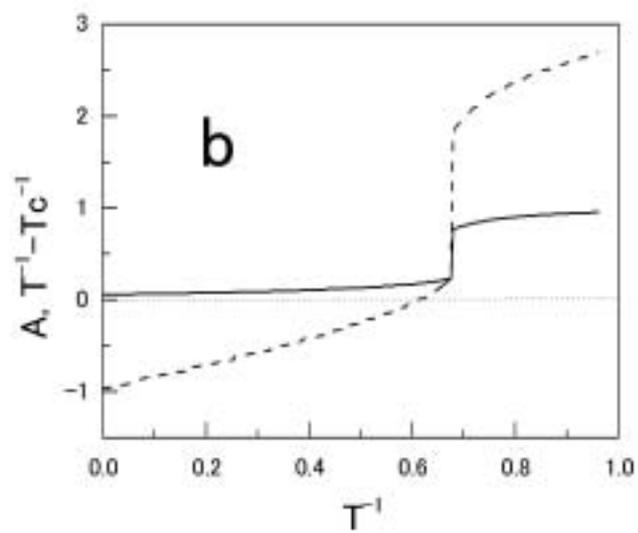

Fig.6 Sasai

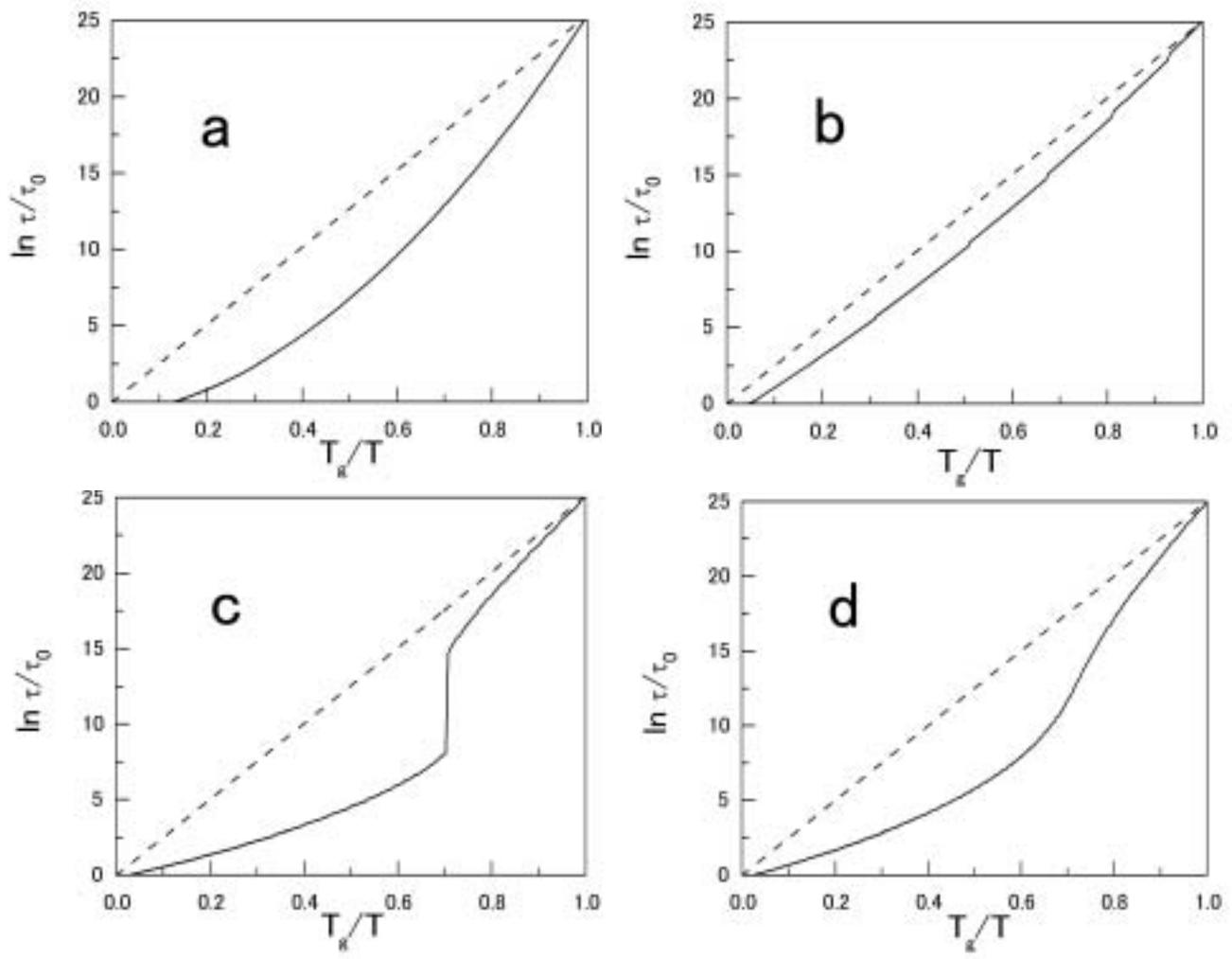

Fig.7 Sasai